\documentclass{iau_FM}
\usepackage{graphicx}

\title[] 
{Characterizing Blue Straggler Star Populations in Globular Clusters using HST Photometric Survey Data}

\author[Mirko Simunovic \& Thomas H. Puzia]   
{Mirko Simunovic$^1$
 \and Thomas H. Puzia$^1$}

\affiliation{$^1$Institute of Astrophysics, Pontificia Universidad Cat\'olica de Chile \\ Avenida Vicu\~na Mackenna 4860,
Santiago, Chile \\ email: {\tt msimunov@astro.puc.cl, tpuzia@astro.puc.cl}} 
\pubyear{2015}
\jname{Astronomy in Focus, Volume 1} 
\editors{Piero Benvenuti, ed.}
\begin{document}

\maketitle

\begin{abstract}
 We present early results from a detailed analysis of the BSS population in Galactic GCs based on HST data.\,Using proper motion cleaning of the color-magnitude diagrams we construct a large catalog of BSSs and study some population properties.\,Stellar evolutionary models are used to find stellar mass and age estimates for the BSS populations in order to establish constraints related to the dynamical interactions in which they may have formed. 
\end{abstract}

\firstsection 
\section{Introduction and Data Description.}

In the denser regions of globular clusters (GC) where two-body relaxation processes are expected to affect global GC properties, close stellar interactions can lead to mass-transfer and/or merger events.\,Blue Straggler Stars (BSS) are likely formed in such interactions, and the cluster structural parameters could be linked to the observed BSS population properties (\cite{leigh13}).\,The cluster dynamical evolution is driving much of the initial conditions, and BSSs are in fact good tracers of their current dynamical state (\cite{ferraro}).\,In this broader context it is that we studied BSSs in a general approach, by looking at them and their properties in multiple galactic GCs.\,This work is based on photometric catalogs obtained from HST archival photometric data in the ACS F606W and F814W filters (PI: A. Sarajedini, Program ID:10775 ), as well as imaging data in the WFC3 F336W filter (PI: G. Piotto, Program ID:13297 ).\,The former catalogs (optical) are obtained directly from the official survey website.\,The latter data (near-UV) has been reduced with standard STScI pipelines.\,Hence, for this work we use multi-epoch and multi-band catalogs for the innermost regions ($\sim$2.7$\times$2.7 arcmin) of 43 GCs.

\section{Analysis and Results.}

We use proper motion measurements to construct clean color-magnitude diagrams (CMD).\,The BSSs are selected from each of the CMDs using the method of \cite{leigh11}.\,Moreover, our decontamination allows us to unambiguously select potential BSSs that have evolved off the main-sequence (MS), and are now evolving through their own SGB phase.\,These Yellow Straggler Stars (YSS) have been easily detected in sparse GCs (\cite{clark}).\,We show in Figure 1a the CMD of NGC\,6717 for illustration of our selection methods. The BSSs and YSSs are plotted as different symbols.\,Overall, we detect YSSs present in the innermost regions of 37 GCs.\,The YSS/BSS ratio (using only BSSs expected to be in the MS) for the vast majority of GCs is between $\sim$10$-$30\%.\,If we assume a roughly constant formation rate in the past few Gyrs, then the MS to SGB lifetime ratio and the ratio of stars observed today at those evolutionary stages are correlated.\,The MS lifetime varies greatly with mass, while the SGB lifetime is much shorter and better constrained.\,Assuming an SGB lifetime of $\sim$200 Myr as representative for the masses expected for BSSs ($\sim$1.0$-$1.2 M$_{\odot}$), the observed YSS to BSS ratio ($\sim$10$-$30\%) suggests a MS lifetime of $\sim$1$-$2 Gyr, which is in very good agreement with \cite{sills}, who proposed similarly a MS lifetime of 1$-$2 Gyr based on the observed frequency of evolved BSSs experiencing a horizontal-branch phase.\,As pointed out by \cite{sills}, this result suggests that BSSs have truncated MS lifetimes for their mass, such that full mixing is unlikely during formation.\,Next, we use a set of Dartmouth isochrones that can encompass the entire BSS region in the CMD and perform a robust interpolation method to find stellar mass and age estimates for each BSS.\,The ones currently in binary systems are expected to be over-luminous therefore adding complexity to the interpretation of our approach. Hence, we limit ourselves to BSSs below the lower luminosity limit expected for BSS in binary systems (\cite{xin}).\,The BSS mass distribution is shown in Figure 1b as a solid line.\,For comparison, we plot in a dashed line the mass distribution obtained by \cite{stepien} using evolutionary models of binary merger products.\,We also plot as a short dashed line the mass distribution obtained by \cite{fiorentino} using pulsation equations from models of nine pulsating BSSs (known as SX Phoenicis stars) observed in NGC\,6541.\,Remarkably, all distributions peak at around $\sim$1.0 M$_{\odot}$, setting a clear constraint for formation models of BSSs.\,We plot in Figure 1c the average BSS mass, normalized to the turn-off mass, as a function of the cluster half-mass relaxation time.\,Considering that we are sampling the innermost regions of GCs, we argue that the plot is suggesting the result of mass segregation affecting BSSs in more dynamically evolved GCs, or that the denser and more compact cores in evolved clusters could be more efficient at making relatively more massive BSSs.\,We note that the former scenario is consistent with the results by \cite{ferraro}.\,This shows the great potential of our approach for providing key empirical relations and thus detailed understanding  of BSS formation channels. We refer the reader to our upcoming paper (Simunovic \etal\ 2015, in prep.) for a detailed description of the analysis.

\begin{figure}[t!]
\begin{center}
 \includegraphics[width=1.74in]{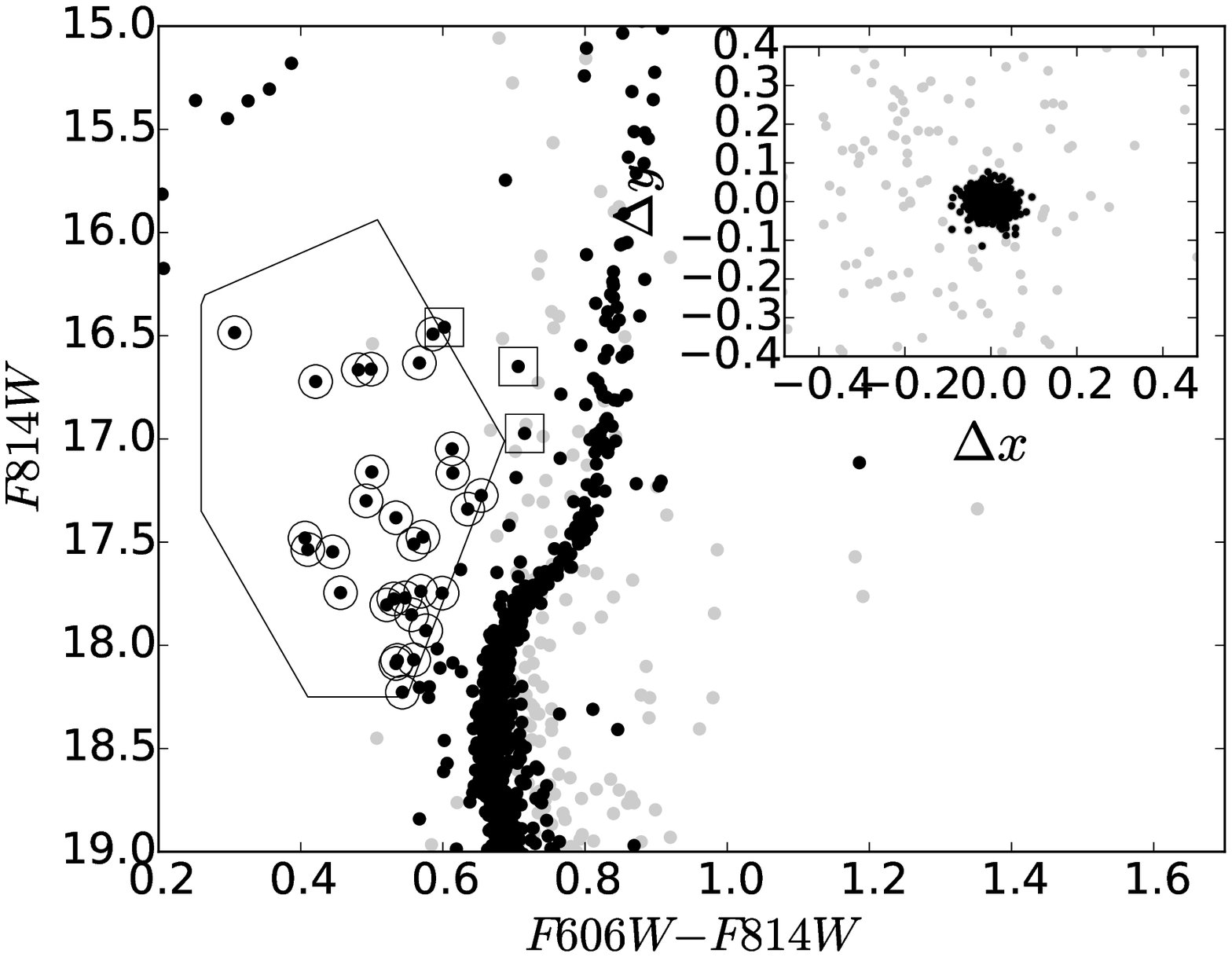} 
 \includegraphics[width=1.74in]{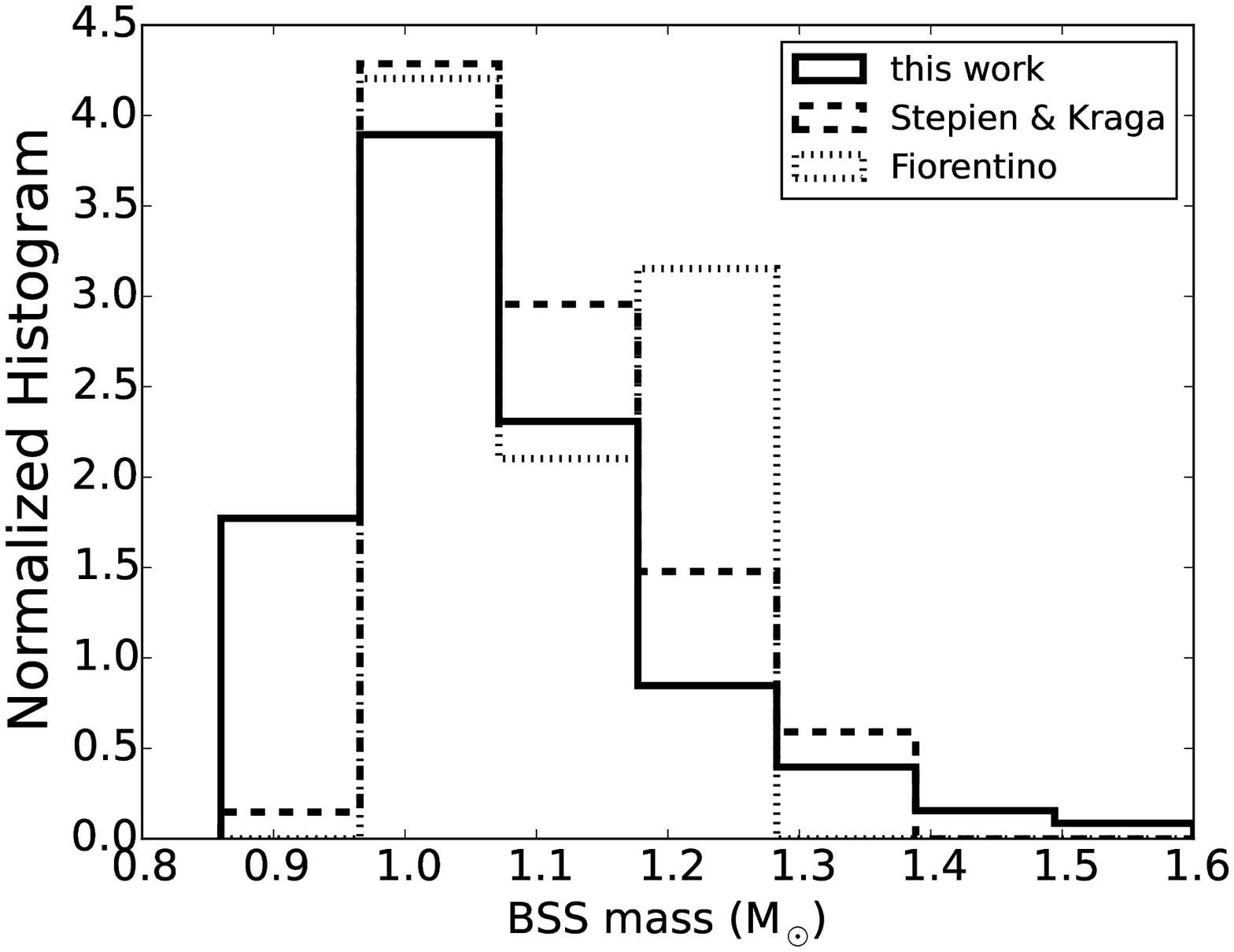} 
 \includegraphics[width=1.74in]{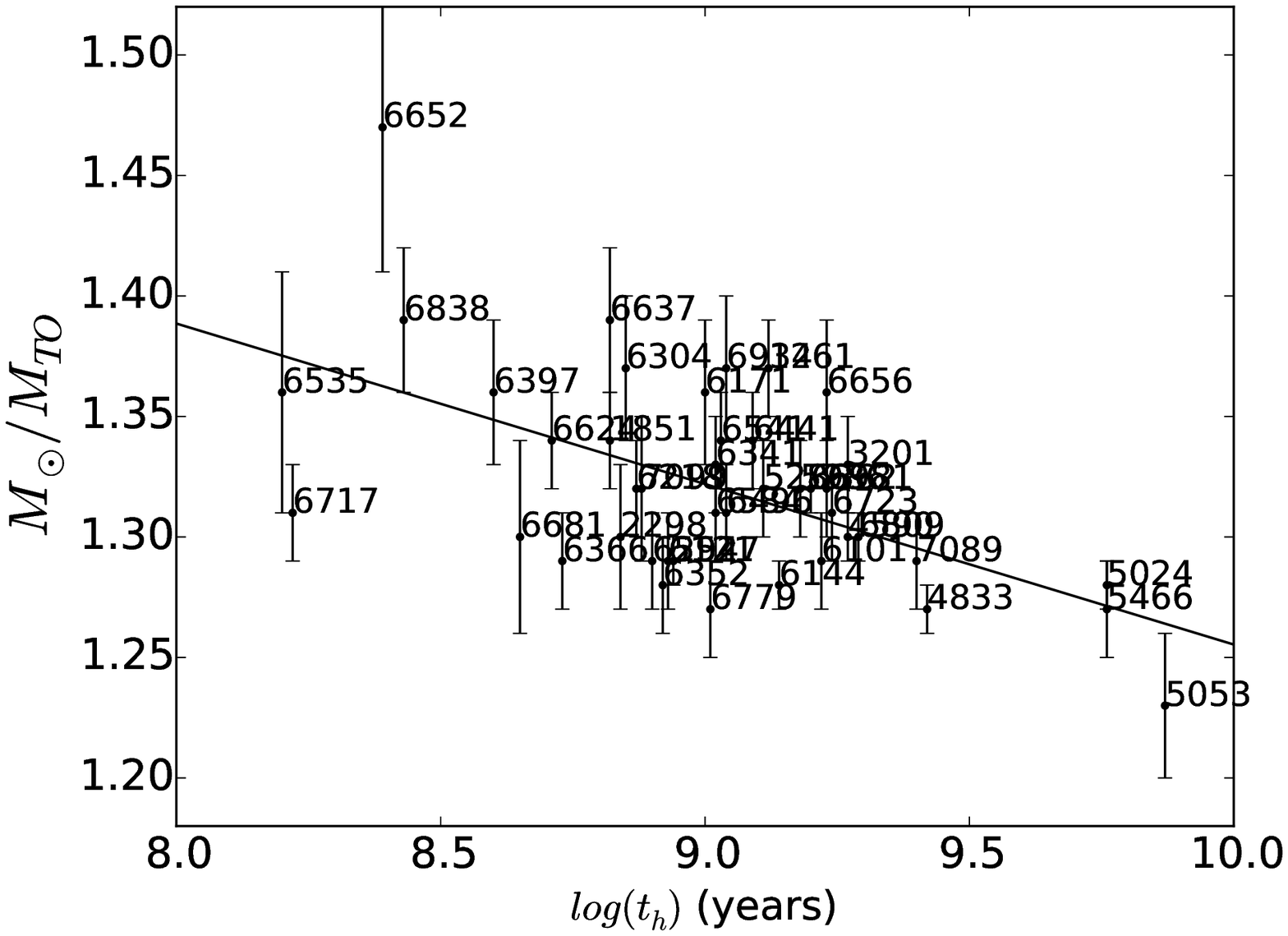} 
 \caption{$a$,$b$ and $c$, from left to right.}
   \label{fig1}
\end{center}
\end{figure}

\end{document}